\documentclass[conference]{IEEEtran}
\usepackage[T1]{fontenc}
\usepackage{lmodern}
\usepackage{amsmath,amssymb,amsfonts}
\usepackage{algorithmic}
\usepackage{graphicx}
\usepackage{textcomp}
\usepackage{xcolor}
\usepackage{booktabs}
\usepackage{multirow}
\usepackage{hyperref}
\usepackage{tabularx}
\usepackage{adjustbox}
\usepackage{caption}
\usepackage{subcaption}
\usepackage{tikz}
\usepackage{pgfplots}
\pgfplotsset{compat=1.18}
\usetikzlibrary{positioning,arrows.meta,shapes.geometric,fit,backgrounds,calc,patterns}
\usepackage{listings}
\def\BibTeX{{\rm B\kern-.05em{\sc i\kern-.025em b}\kern-.08em
    T\kern-.1667em\lower.7ex\hbox{E}\kern-.125emX}}

\begin{document}
\IEEEoverridecommandlockouts
\renewcommand{\footnoterule}{\kern-3pt \hrule width 0.4\columnwidth \kern 2.6pt}

\title{ResRank: Unifying Retrieval and Listwise Reranking via End-to-End Joint Training with Residual Passage Compression}

\author{
\IEEEauthorblockN{Xiaojie Ke\textsuperscript{\dag, *}, Shuai Zhang\textsuperscript{\dag,\ddag}, Liansheng Sun, Yongjin Wang, \\
    Hengjun Jiang, Xiangkun Liu, Cunxin Gu, Jian Xu, Guanjun Jiang}
\IEEEauthorblockA{Qwen Applications Business Group of Alibaba}

\thanks{\textsuperscript{\dag}Equal contribution.}
\thanks{\textsuperscript{\ddag}Work done during an internship at Alibaba.}
\thanks{\textsuperscript{*}Corresponding author: qianyi.kxj@alibaba-inc.com.}
}

\maketitle

% ============================================================
%                       ABSTRACT
% ============================================================
\begin{abstract}
Large language model (LLM) based listwise reranking has emerged as the dominant paradigm for achieving state-of-the-art ranking effectiveness in information retrieval. However, its reliance on feeding full passage texts into the LLM introduces two critical bottlenecks: the ``lost in the middle'' phenomenon degrades ranking quality as input length grows, and the inference latency scales super-linearly with sequence length, rendering it impractical for industrial deployment. In this paper, we present \textbf{ResRank}, a unified retrieval-reranking framework that fundamentally addresses both challenges. Inspired by multimodal LLMs that project visual inputs into compact token representations, ResRank employs an Encoder-LLM to compress each candidate passage into a single embedding, which is then fed alongside the query text into a Reranker-LLM for listwise ranking. To alleviate the misalignment between the compressed representation space and the ranking space, we introduce a residual connection structure that combines encoder embeddings with contextualized hidden states from the reranker. Furthermore, we replace the conventional autoregressive decoding with a one-step cosine-similarity-based scoring mechanism, eliminating the generation bottleneck entirely. ResRank is trained through a carefully designed dual-stage, multi-task, end-to-end joint optimization strategy that simultaneously trains the encoder and reranker, achieving learning objective alignment between retrieval and reranking while substantially reducing training complexity. Extensive experiments on TREC Deep Learning and eight BEIR benchmark datasets demonstrate that ResRank achieves competitive or superior ranking effectiveness compared to existing approaches while requiring zero generated tokens and processing only one token per passage, yielding a fundamentally better balance between effectiveness and efficiency.
\end{abstract}

\begin{IEEEkeywords}
information retrieval, listwise reranking, large language models, passage compression, joint training
\end{IEEEkeywords}

% ============================================================
%                     1. INTRODUCTION
% ============================================================
\section{Introduction}

Modern information retrieval (IR) systems typically employ a multi-stage pipeline in which a lightweight first-stage retriever rapidly recalls candidate passages from a large corpus, followed by a more sophisticated reranker that refines the ranking order~\cite{zhu2024llmir}. With the advent of large language models, the reranking stage has witnessed remarkable progress: LLM-based listwise rerankers~\cite{sun2023rankgpt,pradeep2023rankvicuna} that accept a query together with multiple candidate passages and directly output a permutation have achieved state-of-the-art effectiveness, substantially outperforming traditional cross-encoder approaches.

Despite their effectiveness, deploying LLM-based listwise rerankers at scale faces two fundamental challenges. First, concatenating the full text of dozens or hundreds of candidate passages creates extremely long input sequences. Research has shown that LLMs exhibit a pronounced ``lost in the middle'' phenomenon~\cite{liu2024lost}, where information buried in the middle of long contexts is disproportionately neglected, directly undermining ranking quality. Practitioners have resorted to sliding window strategies~\cite{sun2023rankgpt} to alleviate this issue, but the resulting multi-pass inference multiplies latency by a factor proportional to the number of windows. Second, even when the input length is manageable, the autoregressive decoding process---generating passage identifiers token by token---adds considerable overhead, especially when ranking long candidate lists. Together, these two bottlenecks make standard LLM-based listwise reranking prohibitively expensive for real-time applications.

Recent efforts have sought to mitigate these challenges from different angles. Long-context LLMs have been explored for full-list ranking in a single pass~\cite{liu2024fullrank}, yet the inherent quadratic complexity of self-attention still limits their practicality at scale. PE-Rank~\cite{liu2025perank} proposed compressing each passage into a single embedding before feeding it to the reranker, drawing inspiration from multimodal LLMs that project visual inputs into compact token representations~\cite{liu2023llava}. While this dramatically reduces input length, PE-Rank adopts a two-stage training paradigm---first aligning the encoder's embedding space to the reranker via a separate projection layer, then training the ranking objective---which increases training complexity and precludes a unified optimization of retrieval and reranking goals. Moreover, it still relies on autoregressive constrained decoding for output generation, leaving room for further efficiency improvement. C2R~\cite{zhi2026c2r} extended the compression paradigm to multi-vector surrogates and jointly optimized the compressor and reranker, but it retains autoregressive generation and the sliding window inference strategy, limiting its efficiency gains. E2Rank~\cite{liu2025e2rank} explored using a single embedding model for both retrieval and reranking by reinterpreting the listwise prompt as pseudo-relevance feedback, yet during inference it still requires the full text of all candidate passages as input to construct the enriched query representation, leaving the input length bottleneck fundamentally unresolved.

In this paper, we propose \textbf{ResRank}, a unified framework that comprehensively addresses the efficiency and effectiveness challenges of LLM-based listwise reranking. Our approach is built upon three key technical innovations:

\textbf{(1) Residual passage compression.} Following the multimodal LLM paradigm~\cite{liu2023llava,chen2024hllm}, we employ an Encoder-LLM to compress each candidate passage into a single dense embedding, which is directly fed into the Reranker-LLM's input space. To bridge the gap between the compressed representation space and the ranking-oriented space, we introduce a residual connection that combines the original encoder embedding with the contextualized hidden state produced by the reranker, thereby reducing learning difficulty and preserving passage-level information.

\textbf{(2) Similarity-based scoring.} Instead of relying on costly token-by-token autoregressive decoding, we adopt a retrieval-inspired scoring mechanism: the reranker's hidden state at the end-of-sequence position, enriched by cross-passage contextual attention, serves as a global aggregation embedding, which is directly compared with each passage's fused representation via a single-step cosine similarity computation. This completely eliminates the generation phase, reducing the number of generated tokens to zero.

\textbf{(3) Dual-stage, multi-task, end-to-end joint training.} We jointly optimize the Encoder-LLM and Reranker-LLM in an end-to-end manner through a two-stage supervised fine-tuning process. Multi-task learning is employed to simultaneously optimize ranking and retrieval objectives, ensuring that the encoder preserves its retrieval capability while adapting to the reranker's requirements. This unified training paradigm aligns the learning objectives of both modules, enabling superior end-to-end retrieval-reranking performance.

We conduct comprehensive experiments on TREC Deep Learning 2019 \& 2020~\cite{craswell2020trec19,craswell2021trec20} and eight out-of-domain BEIR~\cite{thakur2021beir} datasets. Results show that ResRank achieves competitive or superior ranking effectiveness compared to strong baselines---including both full-text LLM rerankers and zero-shot approaches powered by GPT-4---while processing only one token per passage and generating zero tokens during inference. Furthermore, end-to-end experiments demonstrate that the jointly trained encoder and reranker achieve aligned learning objectives, yielding the best overall ranking performance when combined with sparse retrieval through reciprocal rank fusion. Thorough ablation studies confirm that every component of our framework---residual connections, dual-stage training, end-to-end optimization, and multi-task learning---contributes indispensably to the final performance.

% ============================================================
%                   2. RELATED WORK
% ============================================================
\section{Related Work}

\subsection{LLM-based Text Reranking}

The integration of large language models into text reranking has given rise to three principal paradigms, each characterized by how candidate passages are presented to and evaluated by the model. \textit{Pointwise} approaches assess each query-passage pair independently, assigning a relevance score or label per document. While conceptually straightforward, early pointwise methods struggled with score calibration across different passages. Guo et al.~\cite{guo2024mcranker} proposed MCRanker, which simulates a multi-perspective annotation team to generate diverse evaluation criteria on the fly, substantially improving pointwise ranking consistency. Cross-encoder architectures such as monoBERT~\cite{nogueira2019monobert} and monoT5~\cite{nogueira2020monot5} also fall into this category, where a pre-trained language model jointly encodes the query and each passage to produce a relevance score.

\textit{Pairwise} methods reduce the ranking problem to a series of binary comparisons between passage pairs. Qin et al.~\cite{qin2023pairwise} demonstrated that pairwise ranking prompting (PRP) achieves strong performance even with moderate-sized open-source LLMs, showing robustness to input ordering that listwise methods lack. However, pairwise approaches inherently require $O(n^2)$ comparisons for full ranking or $O(n \log n)$ with sorting-based strategies, limiting their scalability.

\textit{Listwise} approaches, which present the query and all candidate passages simultaneously and ask the model to produce a complete permutation, have emerged as the most effective paradigm. RankGPT~\cite{sun2023rankgpt} pioneered this direction by leveraging GPT-3.5 and GPT-4 as zero-shot listwise rerankers, achieving remarkable effectiveness without any task-specific training. Subsequent work has focused on distilling this capability into smaller, deployable models: RankVicuna~\cite{pradeep2023rankvicuna} and RankZephyr~\cite{pradeep2023rankzephyr} distilled listwise ranking knowledge from GPT-4 into 7B-parameter open-source LLMs, while ListT5~\cite{yoon2024listt5} adopted a fusion-in-decoder architecture for zero-shot listwise reranking. TourRank~\cite{chen2025tourrank} introduced a tournament-inspired strategy to enhance LLM-based ranking efficiency. More recently, DiffuRank~\cite{liu2026diffurank} explored diffusion language models as an alternative to autoregressive generation for listwise reranking, enabling parallel denoising-based permutation generation. Beyond these text-in, text-out formulations, jina-reranker-v3~\cite{wang2025jina} proposed a ``last but not late'' interaction mechanism that processes all documents within shared context windows and extracts contextual embeddings for similarity-based scoring, pointing toward more efficient listwise architectures.

\subsection{Efficiency in LLM-based Reranking}

The computational cost of LLM-based listwise reranking stems from two sources: the encoding of long concatenated input sequences and the autoregressive generation of output permutations. The dominant mitigation strategy has been the sliding window approach~\cite{sun2023rankgpt}, which processes a fixed-size subset of passages at each step and slides the window across the candidate list. While this bounds the per-step input length, it introduces multi-pass inference overhead that scales linearly with the candidate list size.

Liu et al.~\cite{liu2024fullrank} investigated whether long-context LLMs could eliminate the need for sliding windows by processing all candidates in a single forward pass. Their findings revealed that while fine-tuned full-ranking models can outperform sliding window counterparts, zero-shot full ranking remains significantly weaker, and the quadratic attention complexity still poses scalability concerns. From the input compression perspective, CompLLM~\cite{berton2025compllm} proposed segment-wise soft compression that reduces token counts by independently compressing short text segments into concept embeddings, achieving linear complexity in context length.

PE-Rank~\cite{liu2025perank} introduced a fundamentally different approach by leveraging dense retrieval encoders to compress each passage into a single embedding. Inspired by multimodal LLMs such as LLaVA~\cite{liu2023llava}, a projection layer maps passage embeddings into the reranker's input space, reducing the per-passage token count from hundreds to one. PE-Rank further proposed dynamic-constrained decoding to restrict the output space to passage embeddings rather than vocabulary tokens. However, its decoupled two-stage training procedure---first training the projector for representation alignment, then training the reranker for ranking---increases pipeline complexity, and the reliance on autoregressive constrained decoding limits throughput. C2R~\cite{zhi2026c2r} extended the compression paradigm by representing each passage as a short sequence of multi-vector surrogates (rather than a single embedding), and jointly fine-tuned the compressor and reranker in an end-to-end fashion so that the compression becomes ranking-aware. Despite its strong effectiveness, C2R still generates output permutations autoregressively and employs the sliding window strategy during inference, constraining its efficiency advantage. E2Rank~\cite{liu2025e2rank} reinterpreted the listwise prompt as a pseudo-relevance feedback query, replacing autoregressive decoding entirely with embedding-space cosine similarity. However, constructing the enriched query representation still requires feeding the full text of all top-$K$ candidate passages into the model at inference time, leaving the input length bottleneck unaddressed. Our work builds upon these insights while addressing their respective limitations: ResRank achieves single-embedding compression without a separate alignment stage, eliminates autoregressive decoding via similarity-based scoring, and unifies retrieval and reranking through end-to-end joint training.

\subsection{Unified Retrieval and Reranking}

The vision of unifying retrieval and reranking within a single framework has attracted growing attention. Wang et al.~\cite{wang2023lsm} articulated the ``Large Search Model'' concept, proposing that a single customized LLM could subsume the entire search stack---from query understanding through ranking to answer generation---using natural language prompts to differentiate tasks. While conceptually appealing, realizing this vision requires addressing the tension between the efficiency demands of retrieval and the effectiveness demands of reranking.

In the recommendation domain, analogous unification efforts have yielded promising results. OneRec~\cite{deng2025onerec} demonstrated that a single generative model with session-wise generation can replace the traditional multi-stage cascade in production recommendation systems. UniSearch~\cite{chen2025unisearch}, deployed at Kuaishou, introduced an end-to-end generative search framework that jointly optimizes tokenization and generation, showing significant real-world improvements. HLLM~\cite{chen2024hllm} proposed a hierarchical two-LLM architecture where an Item-LLM compresses item descriptions into single embeddings that are then processed by a User-LLM, demonstrating the effectiveness of compressed representations in recommendation.

In text retrieval, RocketQAv2~\cite{ren2021rocketqav2} pioneered joint training of a dense retriever and a cross-encoder reranker through dynamic listwise distillation, showing that mutual knowledge transfer between the two modules yields superior performance for both. However, RocketQAv2 operated with relatively small encoder models and did not explore the listwise LLM reranking paradigm. Our ResRank framework advances this line of research by jointly training an Encoder-LLM and a Reranker-LLM under the listwise reranking paradigm, achieving alignment between retrieval and reranking learning objectives while maintaining practical inference efficiency.

% ============================================================
%                   3. METHODOLOGY
% ============================================================
\section{Methodology}

\subsection{Overview}

\begin{figure*}[t]
\centering
\includegraphics[width=0.92\textwidth]{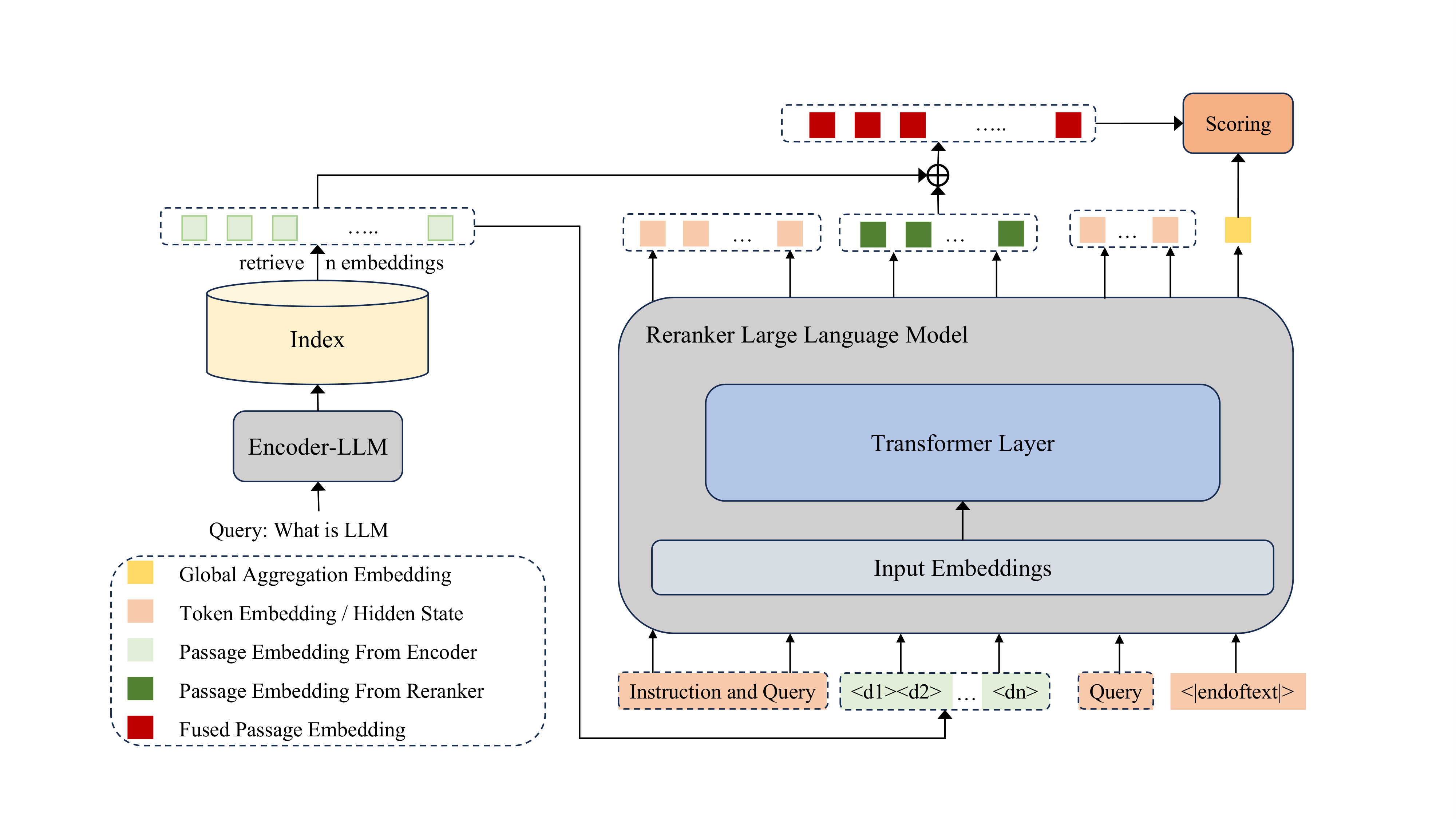}

\caption{Architecture overview of ResRank. Each passage is independently compressed into a single embedding by the Encoder-LLM and directly fed into the Reranker-LLM alongside tokenized instruction and query. The reranker contextualizes all inputs through causal attention, producing hidden states at each passage position. A residual connection combines the original encoder embedding with the reranker's contextualized hidden state to form the fused passage embedding. The global aggregation embedding at the \texttt{[EOS]} position captures the holistic ranking context, and final relevance scores are computed via cosine similarity between the global aggregation embedding and each fused passage embedding, entirely eliminating autoregressive decoding.}
\label{fig:framework}
\end{figure*}

Figure~\ref{fig:framework} illustrates the overall architecture of ResRank. Given a query $q$ and a set of $n$ candidate passages $\mathcal{D} = \{d_1, d_2, \ldots, d_n\}$, our framework operates in three stages: (1) each passage is independently compressed into a single dense embedding by the Encoder-LLM; (2) the passage embeddings are concatenated with the query text and fed into the Reranker-LLM, which contextualizes all inputs through causal attention; and (3) a cosine-similarity-based scoring mechanism produces the final ranking without any autoregressive generation. The entire system, including both the encoder and the reranker, is jointly optimized through a dual-stage, multi-task training procedure.

\subsection{Passage Compression via Encoder-LLM}

The cornerstone of ResRank's efficiency lies in compressing each candidate passage from a variable-length token sequence into a single dense embedding. Formally, let $d_i = (w_1^i, w_2^i, \ldots, w_{L_i}^i)$ denote the token sequence of the $i$-th passage with length $L_i$. The Encoder-LLM $f_\text{enc}$ processes each passage independently and extracts a fixed-dimensional embedding:
\begin{equation}
\mathbf{e}_i = f_\text{enc}(d_i) \in \mathbb{R}^{d},
\end{equation}
where $d$ is the hidden dimension. This embedding encapsulates the semantic content of the entire passage into a compact representation, reducing the per-passage token count from $L_i$ to exactly one.

The choice of the Encoder-LLM is critical, as its embedding quality directly determines the upper bound of information available to the downstream reranker. We adopt Qwen3-Embedding-4B~\cite{zhang2025qwen3emb} as our encoder, a state-of-the-art text embedding model that produces high-quality dense representations suitable for both retrieval and semantic understanding tasks. Crucially, Qwen3-Embedding-4B shares the same hidden dimension and architectural family as the Reranker-LLM (Qwen3-4B~\cite{yang2025qwen3}), which allows the passage embeddings to be \textbf{directly} injected into the reranker's input layer without requiring any intermediate projection or alignment module. This architectural compatibility eliminates the need for a separate projector---a notable simplification over prior approaches such as PE-Rank~\cite{liu2025perank} that require dedicated projection layers to bridge heterogeneous representation spaces.

\subsection{Residual-Enhanced Listwise Reranking}

The passage embeddings from the encoder are treated as special input tokens and concatenated with the tokenized instruction and query text to form the reranker's input:
\begin{equation}
\mathbf{X} = [\mathbf{t}_\text{inst}; \mathbf{t}_q; \mathbf{e}_1, \mathbf{e}_2, \ldots, \mathbf{e}_n; \mathbf{t}_q'; \mathbf{t}_\text{eos}],
\end{equation}
where $\mathbf{t}_\text{inst}$ and $\mathbf{t}_q$ are the token embeddings of the instruction and query, $\mathbf{t}_q'$ denotes an additional query anchor appended after the passage embeddings, and $\mathbf{t}_\text{eos}$ is the end-of-sequence token. The Reranker-LLM $f_\text{rer}$ processes this input through causal self-attention:
\begin{equation}
\mathbf{H} = f_\text{rer}(\mathbf{X}) = [\ldots; \mathbf{h}_1^p, \ldots, \mathbf{h}_n^p; \ldots; \mathbf{h}_\text{eos}],
\end{equation}
where $\mathbf{h}_i^p \in \mathbb{R}^{d}$ denotes the hidden state at the position of the $i$-th passage embedding.

Through the causal attention mechanism, each passage's hidden state $\mathbf{h}_i^p$ is enriched with contextual information from the instruction, query, and all preceding passage embeddings, capturing cross-passage interaction signals that are absent from the original encoder embeddings.

A key observation is that compressing a passage into a single embedding inevitably entails information loss, and the reranker's hidden states alone may not fully compensate for this deficit, especially during the early stages of training when the cross-module alignment is imperfect. To alleviate this and reduce learning difficulty, we introduce a \textbf{residual connection} that combines the original encoder embedding with the reranker's contextualized hidden state to form the \textit{fused passage embedding}:
\begin{equation}
\mathbf{r}_i = \mathbf{h}_i^p + \mathbf{e}_i.
\label{eq:residual}
\end{equation}

This residual structure serves a dual purpose: $\mathbf{e}_i$ preserves the encoder's passage-level semantic information that might be attenuated through the reranker's attention layers, while $\mathbf{h}_i^p$ supplements this with contextualized signals arising from query-passage and cross-passage interactions within the reranker. The additive combination provides a gradient shortcut that stabilizes training and lowers the optimization barrier.

\subsection{Cosine-Similarity-Based Scoring}

Conventional listwise rerankers produce rankings through autoregressive decoding, generating passage identifiers token by token. This process contributes substantial latency: for a list of $n$ passages, a typical generative reranker must produce approximately $n$ to $4.5n$ output tokens~\cite{liu2025perank}. Even with constrained decoding strategies that restrict the output vocabulary, the sequential nature of autoregressive generation remains a bottleneck.

We circumvent this entirely by adopting a retrieval-inspired scoring mechanism. The hidden state $\mathbf{h}_\text{eos}$ at the \texttt{[EOS]} position, having attended to all instruction tokens, query tokens, and passage embeddings through the full depth of the reranker, serves as a \textit{global aggregation embedding} that encapsulates the holistic context of the ranking problem. The relevance score for each passage is then computed via cosine similarity:
\begin{equation}
s(q, d_i) = \frac{\mathbf{h}_\text{eos}^\top \mathbf{r}_i}{\|\mathbf{h}_\text{eos}\| \cdot \|\mathbf{r}_i\|}.
\label{eq:score}
\end{equation}

The final ranking is obtained by sorting passages according to their scores in descending order. This formulation reduces the number of generated tokens from $O(n)$ to exactly \textbf{zero}, transforming the ranking operation from a sequential generation problem into a parallel computation. The entire inference pipeline---encoding, reranking, and scoring---requires only two forward passes (one through the encoder, one through the reranker) with no autoregressive decoding loop.

\subsection{Training Strategy}

\subsubsection{Multi-Task Learning Objective}

We first describe the loss function used for ResRank. For the encoder, we employ standard contrastive learning to align relevant query-document pairs while pushing apart irrelevant ones. Specifically, for a training query $q_i$, there is one positive document $d_i^+$ and a set of negative documents $D^-$. Given a batch of $N$ instances, we minimize the InfoNCE loss:
\begin{equation}
\small
    \mathcal{L}_{\text{InfoNCE}} = - \frac{1}{N} \sum_{i=1}^N \log \frac{e^{s_{\text{enc}}(q_i, d_i^+)/\tau_1}}{\displaystyle e^{s_{\text{enc}}(q_i, d_i^+)/\tau_1} + \!\sum_{d_j\in D^-}\! e^{s_{\text{enc}}(q_i, d_j)/\tau_1}},
\end{equation}
where $\tau_1$ is a temperature hyperparameter set to 0.05, and $s_{\text{enc}}$ denotes the cosine similarity between the query and document embeddings generated by the encoder.

For the reranker, we optimize the RankNet loss~\cite{burges2005ranknet} to measure the correctness of relative orderings. The relevance score is computed as the cosine similarity between the fused passage embedding $\mathbf{r}_i$ and the global aggregation embedding $\mathbf{h}_\text{eos}$, as defined in Eq.~\eqref{eq:score}. Let $S(q_i, d_j) = \cos(\mathbf{r}_{j}, \mathbf{h}_{\text{eos}}^i)$. The RankNet loss is:
\begin{equation}
    \mathcal{L}_{\text{RankNet}} = \frac{1}{N} \sum_{i=1}^N \sum_{\substack{d_j, d_k \in D \\ r_j < r_k}} \log\left(1 + e^{(S(q_i, d_k) - S(q_i, d_j))/\tau_2}\right),
\end{equation}
where $D$ includes both positive and negative documents, $\tau_2 = 0.05$ scales the similarity scores, and $r_j$ denotes the rank of document $d_j$ (smaller values indicate higher relevance). The indicator $r_j < r_k$ ensures the loss is computed only for pairs where $d_j$ should be ranked above $d_k$.

The final training objective combines both losses:
\begin{equation}
    \mathcal{L} = \lambda \, \mathcal{L}_{\text{InfoNCE}} + \mathcal{L}_{\text{RankNet}},
\label{eq:total_loss}
\end{equation}
where $\lambda = 0.1$ balances the two tasks. This multi-task design is crucial: without the retrieval auxiliary loss, the encoder's standalone retrieval capability degrades severely (as we demonstrate in Section~\ref{sec:ablation}), precluding its use in the first-stage retrieval of an end-to-end pipeline.

\subsubsection{Dual-Stage Supervised Fine-Tuning}

We adopt a two-stage SFT process that transitions from coarse-grained alignment to fine-grained ranking refinement.

\textbf{Stage 1: Coarse-grained alignment.} The first stage trains on a larger set of samples to establish the fundamental alignment between the encoder's compressed representations and the reranker's ranking behavior, enabling the model to learn the basic mechanics of ranking from passage embeddings.

\textbf{Stage 2: Fine-grained refinement.} The second stage introduces a smaller set of high-quality samples, each consisting of one query, one positive document, and multiple negative documents, providing fine-grained contrastive signals. This stage refines the model's ability to distinguish subtle relevance differences among competitive candidates.

Both stages optimize the combined objective in Eq.~\eqref{eq:total_loss}, ensuring that the encoder's retrieval capability is preserved throughout the entire training process.

\subsubsection{End-to-End Joint Optimization}

Unlike approaches that decouple the training of the encoder and reranker into separate stages---for example, PE-Rank~\cite{liu2025perank} first trains a projector for alignment and then fine-tunes the reranker for ranking---ResRank performs full-parameter end-to-end joint optimization of both the Encoder-LLM and the Reranker-LLM from the outset. This design choice is motivated by two considerations. First, decoupled training restricts the passage embeddings to a representation space that may not be optimally suited for the downstream reranking task. Second, joint training enables the encoder to adapt its representations in response to the reranker's feedback signals, while the reranker simultaneously learns to better leverage the encoder's evolving outputs. This mutual adaptation creates a synergistic training dynamic that yields better-aligned representations across the two modules.

% ============================================================
%                   4. EXPERIMENTS
% ============================================================
\section{Experiments}

\subsection{Experimental Setup}

\subsubsection{Datasets and Metrics}

We evaluate ResRank on both in-domain and out-of-domain benchmarks. For in-domain evaluation, we use the TREC Deep Learning 2019~\cite{craswell2020trec19} and 2020~\cite{craswell2021trec20} test sets. For out-of-domain evaluation, we report results on eight datasets from the BEIR benchmark~\cite{thakur2021beir}: Covid, NFCorpus, Touche, DBPedia, SciFact, Signal, News, and Robust. All models rerank the top-100 passages retrieved by BM25. We adopt nDCG@10 as the primary evaluation metric, following standard practice in the field.

\subsubsection{Training Data}

Our training strategy consists of a two-stage supervised fine-tuning process. The first stage trains on 232,419 samples derived from PE-Rank~\cite{liu2025perank} and re-annotated using Qwen3-Max to ensure consistent and high-quality relevance judgments. The second stage trains on approximately 87,000 high-quality samples sourced from E2Rank~\cite{liu2025e2rank}, also re-annotated via Qwen3-Max. Each second-stage sample consists of one query, one positive passage, and 15 negative passages. The annotation prompt is detailed in Appendix~\ref{sec:annotation_prompt}.

\subsubsection{Baselines}

We compare ResRank against a comprehensive suite of reranking models spanning four categories:

\begin{itemize}
\item \textbf{Cross-encoders:} monoBERT~\cite{nogueira2019monobert} and monoT5~\cite{nogueira2020monot5}, representing supervised models trained with human relevance annotations.
\item \textbf{Zero-shot LLM rerankers:} RankGPT (GPT-3.5 and GPT-4)~\cite{sun2023rankgpt} and TourRank~\cite{chen2025tourrank}, which leverage the inherent ranking capabilities of large language models without task-specific fine-tuning.
\item \textbf{Full-text fine-tuned LLM rerankers:} ListT5~\cite{yoon2024listt5}, RankVicuna~\cite{pradeep2023rankvicuna}, RankZephyr~\cite{pradeep2023rankzephyr}, and RankMistral~\cite{liu2025perank}, which fine-tune LLMs on ranking data using full passage text.
\item \textbf{Compressed-token fine-tuned LLM rerankers:} PE-Rank~\cite{liu2025perank}, which compresses each passage into a single embedding for efficient listwise reranking.
\end{itemize}

\subsubsection{Implementation Details}

We employ Qwen3-4B~\cite{yang2025qwen3} as the Reranker-LLM, chosen for its strong instruction-following capability, and Qwen3-Embedding-4B~\cite{zhang2025qwen3emb} as the Encoder-LLM. Both models are jointly optimized with full-parameter training. The first stage trains for one epoch with an effective batch size of 128 and a learning rate of $6 \times 10^{-6}$. The second stage continues for one additional epoch with identical hyperparameters. All experiments are implemented in PyTorch with FlashAttention~\cite{dao2022flashattention} for efficient attention computation and DeepSpeed ZeRO-2~\cite{rasley2020deepspeed} for distributed training and memory optimization.

For fair comparison with baseline models, we adopt the sliding window evaluation protocol (denoted ResRank$_\text{sw}$) in the main comparison experiments, consistent with the evaluation methodology of prior work. However, we emphasize that in practical industrial search engines, the sliding window strategy entails multiplicative latency overhead and is challenging to deploy. Therefore, in all other experiments, unless explicitly stated, we evaluate ResRank in a \textit{single-pass} mode that ranks all candidates simultaneously, which represents its intended deployment configuration.

\subsection{Main Results}

\subsubsection{Out-of-Domain Evaluation (BEIR)}

Table~\ref{tab:beir} presents the reranking results on eight BEIR datasets. Several key observations emerge from this comparison.

\begin{table*}[t]
\centering
\caption{Results (nDCG@10) of reranking top-100 passages on the BEIR benchmark. \textit{Ret.} denotes the first-stage retrieval model. The overall best result is in \textbf{bold}; the best within each block is \underline{underlined}.}
\label{tab:beir}
\resizebox{\textwidth}{!}{
\begin{tabular}{l|c|cccccccc|c}
\toprule
\textbf{Model} & \textbf{Ret.} & \textbf{Covid} & \textbf{NFCorpus} & \textbf{Touche} & \textbf{DBPedia} & \textbf{SciFact} & \textbf{Signal} & \textbf{News} & \textbf{Robust} & \textbf{Avg.} \\
\midrule
BM25 & -- & 0.5947 & 0.3375 & \textbf{0.4422} & 0.3180 & 0.6789 & 0.3305 & 0.3952 & 0.4070 & 0.4380 \\
\midrule
\multicolumn{11}{l}{\textit{Supervised models trained with human annotations}} \\
monoBERT & BM25 & 0.7001 & 0.3688 & 0.3175 & 0.4187 & 0.7136 & 0.3144 & 0.4462 & 0.4935 & 0.4716 \\
monoT5 & BM25 & \underline{0.8071} & \underline{0.3897} & \underline{0.3241} & \underline{0.4445} & \underline{0.7657} & \underline{0.3255} & \underline{0.4849} & \underline{0.5671} & \underline{0.5136} \\
\midrule
\multicolumn{11}{l}{\textit{Unsupervised LLM-based listwise models}} \\
RankGPT-3.5 & & 0.7667 & 0.3562 & 0.3618 & 0.4447 & 0.7043 & 0.3212 & 0.4885 & 0.5062 & 0.4937 \\
RankGPT-4 & 3$\times$BM25 & \underline{\textbf{0.8551}} & \underline{0.3847} & \underline{0.3857} & \underline{\textbf{0.4712}} & \underline{0.7495} & \underline{0.3440} & \underline{0.5289} & 0.5755 & \underline{0.5368} \\
TourRank & & 0.8259 & 0.3799 & 0.2998 & 0.4464 & 0.7217 & 0.3083 & 0.5146 & \underline{0.5787} & 0.5094 \\
\midrule
\multicolumn{11}{l}{\textit{LLM-based listwise models trained with distillation}} \\
RankMistral & & 0.7800 & 0.3310 & 0.2746 & 0.3771 & 0.6622 & 0.3004 & 0.3710 & 0.3954 & 0.4365 \\
ListT5-base & & 0.7830 & 0.3560 & 0.3340 & 0.4370 & 0.7410 & 0.3350 & 0.4850 & 0.5210 & 0.5090 \\
ListT5-3B & 6$\times$BM25 & 0.8470 & 0.3770 & 0.3360 & \underline{0.4620} & \underline{\textbf{0.7700}} & 0.3380 & 0.5320 & 0.5780 & 0.5300 \\
PE-Rank & & 0.7772 & 0.3639 & 0.3306 & 0.4005 & 0.6938 & 0.3374 & 0.4970 & 0.4740 & 0.4843 \\
ResRank & & 0.8409 & 0.3973 & 0.3948 & 0.4583 & 0.7642 & \underline{\textbf{0.3527}} & \underline{\textbf{0.5485}} & 0.5964 & 0.5440 \\
ResRank$_\text{sw}$ & & \underline{0.8500} & \underline{\textbf{0.3994}} & \underline{0.3994} & 0.4547 & 0.7696 & 0.3337 & 0.5332 & \underline{\textbf{0.6182}} & \underline{\textbf{0.5448}} \\
\bottomrule
\end{tabular}
}
\end{table*}

First, ResRank achieves the highest average nDCG@10 among all distillation-trained baselines in single-pass mode, substantially outperforming RankMistral, ListT5-3B, and PE-Rank. Notably, the improvement over PE-Rank---the most directly comparable baseline that also operates on compressed passage embeddings---exceeds 5 absolute points, demonstrating the combined benefits of residual connections, similarity-based scoring, and end-to-end joint training.

Second, ResRank in single-pass mode even surpasses several zero-shot LLM rerankers including RankGPT-3.5 and TourRank, and approaches RankGPT-4 which relies on one of the most powerful commercial LLMs. On specific datasets, ResRank achieves the best overall results on Signal and News.

Third, when evaluated with the sliding window protocol (ResRank$_\text{sw}$), ResRank reaches the highest average nDCG@10 of 0.5448, surpassing all baselines including RankGPT-4. ResRank$_\text{sw}$ also achieves the best scores on NFCorpus and Robust.

\subsubsection{In-Domain Evaluation (TREC DL)}

Table~\ref{tab:trec} reports results on the TREC Deep Learning 2019 and 2020 benchmarks. In single-pass mode, ResRank outperforms PE-Rank by notable margins on both DL19 and DL20, and surpasses most distilled LLM rerankers. With the sliding window protocol, ResRank$_\text{sw}$ achieves the second-best result on DL19 among all methods, trailing only RankGPT-4. On DL20, ResRank$_\text{sw}$ delivers performance on par with distilled full-text rerankers such as ListT5-3B, while operating on compressed single-token representations rather than full passage text---a fundamentally different efficiency regime.

\begin{table}[t]
\centering
\caption{Results (nDCG@10) of reranking top-100 passages on TREC DL. \textit{Ret.} denotes the first-stage retrieval model. Best overall in \textbf{bold}; best per block \underline{underlined}.}
\label{tab:trec}
\begin{tabular}{l|c|cc}
\toprule
\textbf{Model} & \textbf{Ret.} & \textbf{DL19} & \textbf{DL20} \\
\midrule
BM25 & -- & 0.5058 & 0.4796 \\
\midrule
\multicolumn{4}{l}{\textit{Supervised (human annotations)}} \\
monoBERT & 2$\times$BM25 & 0.7050 & 0.6728 \\
monoT5 & & \underline{0.7183} & \underline{0.6889} \\
\midrule
\multicolumn{4}{l}{\textit{Unsupervised LLM-based listwise}} \\
RankGPT-3.5 & & 0.6580 & 0.6291 \\
RankGPT-4 & 3$\times$BM25 & \underline{\textbf{0.7559}} & \underline{0.7056} \\
TourRank & & 0.7163 & 0.6956 \\
\midrule
\multicolumn{4}{l}{\textit{LLM-based listwise (distillation)}} \\
RankVicuna & & 0.6682 & 0.6549 \\
RankZephyr & & 0.7420 & \underline{\textbf{0.7086}} \\
RankMistral & & 0.7173 & 0.6807 \\
ListT5-base & 8$\times$BM25 & 0.7180 & 0.6810 \\
ListT5-3B & & 0.7180 & 0.6910 \\
PE-Rank & & 0.7048 & 0.6354 \\
ResRank & & 0.7359 & 0.6865 \\
ResRank$_\text{sw}$ & & \underline{0.7457} & 0.6906 \\
\bottomrule
\end{tabular}
\end{table}

\subsection{End-to-End Retrieval-Reranking Performance}

A distinctive advantage of ResRank's joint training paradigm is that the encoder, optimized alongside the reranker with a retrieval auxiliary loss, can simultaneously serve as the first-stage retriever in a unified pipeline. Table~\ref{tab:e2e} evaluates this end-to-end capability. Unlike the previous tables where all methods rerank the same BM25-retrieved top-100 candidates, here the first-stage retrieval source varies---BM25 alone, the ResRank encoder alone, or a fusion of both via reciprocal rank fusion (RRF)~\cite{cormack2009rrf} with $K=60$---resulting in different candidate sets for reranking. The reported nDCG@10 scores therefore reflect the combined effect of retrieval recall and reranking precision.

\begin{table*}[t]
\centering
\caption{End-to-end retrieval-reranking results (nDCG@10) on BEIR and TREC DL benchmarks. Each row uses a different first-stage retrieval method (\textit{Ret.}) to produce the top-100 candidate set, which is then reranked by ResRank. The reported metrics reflect the combined effect of retrieval quality and reranking quality. Best results in \textbf{bold}.}
\label{tab:e2e}
\resizebox{\textwidth}{!}{
\begin{tabular}{l|c|cccccccc|c|cc}
\toprule
\textbf{Model} & \textbf{Ret.} & \textbf{Covid} & \textbf{NFCorpus} & \textbf{Touche} & \textbf{DBPedia} & \textbf{SciFact} & \textbf{Signal} & \textbf{News} & \textbf{Robust} & \textbf{BEIR Avg.} & \textbf{DL19} & \textbf{DL20} \\
\midrule
 & BM25 & 0.8409 & 0.3973 & \textbf{0.3948} & 0.4583 & 0.7642 & \textbf{0.3527} & \textbf{0.5485} & 0.5964 & 0.5440 & 0.7359 & 0.6865 \\
ResRank & ResRank & 0.8643 & 0.4313 & 0.3245 & 0.4885 & \textbf{0.7941} & 0.2773 & 0.4113 & 0.6626 & 0.5317 & 0.7634 & 0.7462 \\
 & ResRank+BM25 & \textbf{0.8839} & \textbf{0.4342} & 0.3626 & \textbf{0.5038} & 0.7809 & 0.3094 & 0.4698 & \textbf{0.6771} & \textbf{0.5527} & \textbf{0.7821} & \textbf{0.7499} \\
\bottomrule
\end{tabular}
}
\end{table*}

Several findings are noteworthy. When ResRank's own encoder serves as the retriever, the end-to-end pipeline achieves substantially stronger performance on the TREC benchmarks compared to BM25-based retrieval, demonstrating that the jointly trained encoder provides candidates that are better aligned with the reranker's ranking behavior. On BEIR, the ResRank encoder excels on datasets where dense retrieval has inherent advantages (e.g., NFCorpus, Robust, SciFact), while BM25 remains stronger on lexically-oriented datasets (e.g., Signal, News, Touche).

Recognizing the complementary strengths of sparse and dense retrieval, we fuse both signals through RRF. The resulting ``ResRank+BM25'' configuration achieves the best performance across virtually all settings, obtaining the highest average BEIR score and the best TREC DL scores. These results confirm that end-to-end joint training successfully aligns the retrieval and reranking learning objectives: the encoder produces candidate sets that the reranker can most effectively rank, and the fusion of sparse and dense signals further amplifies this advantage.

\subsection{Inference Efficiency Analysis}

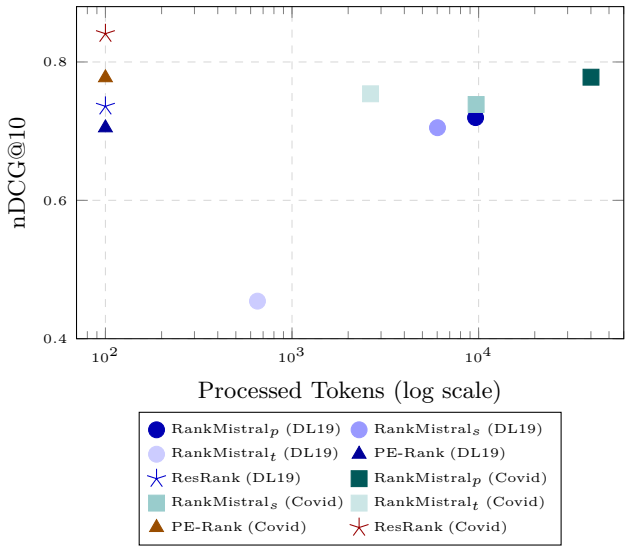
\begin{figure}[t]
\centering
\begin{tikzpicture}
\begin{axis}[
    width=\columnwidth,
    height=6cm,
    xlabel={\small Processed Tokens (log scale)},
    ylabel={\small nDCG@10},
    xmode=log,
    xmin=70, xmax=60000,
    ymin=0.40, ymax=0.88,
    legend style={at={(0.5,-0.22)}, anchor=north, font=\tiny, cells={anchor=west}, legend columns=2},
    grid=major,
    grid style={dashed, gray!30},
    every axis label/.style={font=\small},
    tick label style={font=\tiny},
]

% DL19 data points
\addplot[only marks, mark=*, mark size=3pt, blue!70!black] coordinates {(9635, 0.7196)};
\addplot[only marks, mark=*, mark size=3pt, blue!40] coordinates {(6021, 0.7050)};
\addplot[only marks, mark=*, mark size=3pt, blue!20] coordinates {(653, 0.4543)};
\addplot[only marks, mark=triangle*, mark size=3pt, blue!60!black] coordinates {(100, 0.7048)};
\addplot[only marks, mark=star, mark size=3.75pt, blue!80!black] coordinates {(100, 0.7359)};

% Covid data points
\addplot[only marks, mark=square*, mark size=3pt, teal!70!black] coordinates {(40039, 0.7780)};
\addplot[only marks, mark=square*, mark size=3pt, teal!40] coordinates {(9702, 0.7385)};
\addplot[only marks, mark=square*, mark size=3pt, teal!20] coordinates {(2636, 0.7540)};
\addplot[only marks, mark=triangle*, mark size=3pt, orange!60!black] coordinates {(100, 0.7772)};
\addplot[only marks, mark=star, mark size=3.75pt, red!60!black] coordinates {(100, 0.8409)};

\legend{
    RankMistral$_p$ (DL19),
    RankMistral$_s$ (DL19),
    RankMistral$_t$ (DL19),
    PE-Rank (DL19),
    ResRank (DL19),
    RankMistral$_p$ (Covid),
    RankMistral$_s$ (Covid),
    RankMistral$_t$ (Covid),
    PE-Rank (Covid),
    ResRank (Covid),
}

\end{axis}
\end{tikzpicture}
\caption{Effectiveness vs.\ efficiency trade-off on TREC DL19 and Covid. ResRank achieves the highest nDCG@10 while processing only 100 tokens (one per passage) and generating zero output tokens.}
\label{fig:efficiency}
\end{figure}

Table~\ref{tab:efficiency} and Figure~\ref{fig:efficiency} present a detailed efficiency comparison for reranking the top-100 candidates. We analyze four dimensions: the ranking effectiveness (nDCG@10), the total number of processed tokens (\#Proc.), the average token length per passage (Avg.~$L_p$), and the number of generated tokens (\#Gen.).

\begin{table}[t]
\centering
\caption{Efficiency analysis for reranking top-100 candidates retrieved by BM25. Subscripts of RankMistral indicate input form: original passage (p), summary (s), or title (t).}
\label{tab:efficiency}
\resizebox{\columnwidth}{!}{
\begin{tabular}{l|cccc|cccc}
\toprule
\multirow{2}{*}{\textbf{Model}} & \multicolumn{4}{c|}{\textbf{TREC DL19}} & \multicolumn{4}{c}{\textbf{Covid}} \\
& nDCG & \#Proc. & Avg.~$L_p$ & \#Gen. & nDCG & \#Proc. & Avg.~$L_p$ & \#Gen. \\
\midrule
RankMistral$_\text{p}$ & .7196 & 9635 & 96.4 & 910 & .7780 & 40039 & 400.4 & 987 \\
RankMistral$_\text{s}$ & .7050 & 6021 & 60.2 & 882 & .7385 & 9702 & 97.0 & 930 \\
RankMistral$_\text{t}$ & .4543 & 653 & 6.5 & 865 & .7540 & 2636 & 26.4 & 917 \\
PE-Rank & .7048 & \textbf{100} & \textbf{1.0} & 180 & .7772 & \textbf{100} & \textbf{1.0} & 180 \\
ResRank & \textbf{.7359} & \textbf{100} & \textbf{1.0} & \textbf{0} & \textbf{.8409} & \textbf{100} & \textbf{1.0} & \textbf{0} \\
\bottomrule
\end{tabular}
}
\end{table}

The analysis reveals a striking efficiency advantage. RankMistral, operating on full passage text, must process thousands to tens of thousands of tokens for the concatenated candidate passages, with the enormous variation stemming from differences in average passage length across datasets. It further generates approximately 900 output tokens through autoregressive decoding. When passages are compressed via summarization (RankMistral$_\text{s}$) or truncated to titles (RankMistral$_\text{t}$), the processed token count decreases but effectiveness also drops sharply---particularly RankMistral$_\text{t}$, whose DL19 nDCG@10 collapses dramatically, underscoring that naive text compression sacrifices critical semantic information.

Both PE-Rank and ResRank reduce the per-passage token count to exactly 1.0, yielding a fixed total of 100 processed tokens regardless of the dataset's passage length distribution. This represents up to two orders of magnitude reduction compared to full-text RankMistral. The critical difference between PE-Rank and ResRank lies in the generation phase: PE-Rank still generates 180 tokens through its constrained decoding mechanism, while ResRank generates \textbf{zero} tokens by replacing autoregressive decoding with cosine-similarity-based scoring. Despite this additional efficiency gain, ResRank achieves substantially higher nDCG@10 on both datasets, confirming that the effectiveness-efficiency trade-off is not merely preserved but fundamentally improved.

\subsection{Ablation Studies}
\label{sec:ablation}

To validate the contribution of each component in ResRank, we conduct systematic ablation experiments, with results reported in Table~\ref{tab:ablation}.

\begin{table*}[t]
\centering
\caption{Ablation study results (nDCG@10). Each row removes one component from the full ResRank model.}
\label{tab:ablation}
\resizebox{\textwidth}{!}{
\begin{tabular}{l|cccccccc|c|cc}
\toprule
\textbf{Model} & \textbf{Covid} & \textbf{NFCorpus} & \textbf{Touche} & \textbf{DBPedia} & \textbf{SciFact} & \textbf{Signal} & \textbf{News} & \textbf{Robust} & \textbf{BEIR Avg.} & \textbf{DL19} & \textbf{DL20} \\
\midrule
ResRank (full) & 0.8409 & \textbf{0.3973} & 0.3948 & 0.4583 & 0.7642 & 0.3527 & \textbf{0.5485} & 0.5964 & 0.5440 & \textbf{0.7359} & \textbf{0.6865} \\
\midrule
W/O 1st Stage & 0.8249 & 0.3948 & \textbf{0.4366} & 0.4436 & \textbf{0.7694} & 0.3523 & 0.5366 & 0.5903 & 0.5436 & 0.7091 & 0.6295 \\
W/O 2nd Stage & 0.8469 & 0.3940 & 0.3937 & 0.4421 & 0.7551 & 0.3534 & 0.5365 & 0.5875 & 0.5385 & 0.7210 & 0.6791 \\
\midrule
W/O Hidden State & \textbf{0.8528} & 0.3883 & 0.2850 & 0.4338 & 0.7300 & 0.2877 & 0.5180 & 0.5771 & 0.5091 & 0.7245 & 0.6585 \\
W/O Residual Conn. & 0.8293 & 0.3871 & 0.4314 & 0.4486 & 0.6960 & 0.3227 & 0.5350 & 0.5936 & 0.5305 & 0.7250 & 0.6652 \\
\midrule
W/O Encoder SFT & 0.8327 & 0.3920 & 0.4268 & 0.4496 & 0.7482 & 0.3598 & 0.5324 & 0.5927 & 0.5418 & 0.7148 & 0.6633 \\
W/O Encoder Loss & 0.8406 & 0.3960 & 0.4072 & \textbf{0.4589} & 0.7643 & \textbf{0.3610} & 0.5419 & \textbf{0.5995} & \textbf{0.5462} & 0.7352 & 0.6791 \\
\bottomrule
\end{tabular}
}
\end{table*}

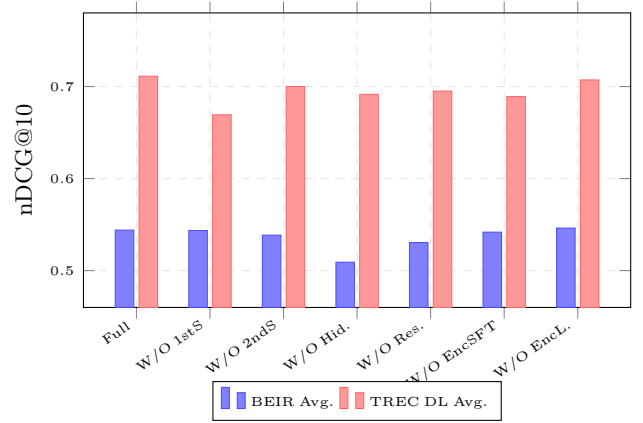
\begin{figure}[t]
\centering
\begin{tikzpicture}
\begin{axis}[
    ybar,
    width=\columnwidth,
    height=5.5cm,
    bar width=7pt,
    ylabel={\small nDCG@10},
    symbolic x coords={Full, {W/O 1stS}, {W/O 2ndS}, {W/O Hid.}, {W/O Res.}, {W/O EncSFT}, {W/O EncL.}},
    xtick=data,
    x tick label style={rotate=35, anchor=east, font=\tiny},
    ymin=0.46, ymax=0.78,
    legend style={at={(0.5,-0.25)}, anchor=north, font=\tiny, legend columns=2},
    every axis label/.style={font=\small},
    tick label style={font=\tiny},
    grid=major,
    grid style={dashed, gray!20},
    enlarge x limits=0.12,
]

\addplot[fill=blue!50, draw=blue!70] coordinates {
    (Full, 0.5440) ({W/O 1stS}, 0.5436) ({W/O 2ndS}, 0.5385) ({W/O Hid.}, 0.5091) ({W/O Res.}, 0.5305) ({W/O EncSFT}, 0.5418) ({W/O EncL.}, 0.5462)
};

\addplot[fill=red!40, draw=red!60] coordinates {
    (Full, 0.7112) ({W/O 1stS}, 0.6693) ({W/O 2ndS}, 0.7001) ({W/O Hid.}, 0.6915) ({W/O Res.}, 0.6951) ({W/O EncSFT}, 0.6891) ({W/O EncL.}, 0.7072)
};

\legend{BEIR Avg., TREC DL Avg.}

\end{axis}
\end{tikzpicture}
\caption{Ablation study summary. BEIR Avg.\ and TREC DL Avg.\ (mean of DL19 and DL20) for the full model and each ablation variant. Every component contributes to the final performance.}
\label{fig:ablation}
\end{figure}

\textbf{Dual-stage training.} Removing the first stage (W/O 1st Stage) causes dramatic degradation on in-domain benchmarks, particularly on DL20 which drops by over 5 points, while the BEIR average remains nearly unchanged. This indicates that the coarse-grained alignment stage is essential for establishing robust in-domain ranking behavior. Conversely, removing the second stage (W/O 2nd Stage) leads to modest but consistent drops across nearly all datasets, demonstrating that fine-grained refinement is necessary to sharpen the model's discriminative capacity. Both stages contribute complementary value.

\textbf{Residual connection structure.} This design is evaluated through two complementary ablations. W/O Hidden State uses only the encoder embedding $\mathbf{e}_i$ as the passage representation, discarding the reranker's contextualized hidden state $\mathbf{h}_i^p$. This variant suffers a severe drop in BEIR average (over 3 points), with particularly large declines on Touche, Signal, and SciFact. Without the hidden states, the model loses all cross-passage contextual signals that the reranker provides, effectively reducing the reranking to a pure embedding similarity computation. Conversely, W/O Residual Connection uses only the reranker's hidden state $\mathbf{h}_i^p$, discarding the skip connection from $\mathbf{e}_i$. This also results in meaningful degradation across most benchmarks. These results confirm that the residual structure successfully combines the encoder's passage-level information with the reranker's contextual enrichment, and that both components are indispensable.

\textbf{End-to-end joint training.} W/O Encoder SFT freezes the encoder during training. This results in decreased performance across both TREC DL and BEIR, confirming that allowing the encoder to adapt alongside the reranker produces better-aligned representations.

\textbf{Multi-task learning.} W/O Encoder Loss removes the retrieval auxiliary objective. Interestingly, the reranking performance remains largely intact, with BEIR average even marginally higher than the full model. However, this ablation has a critical practical consequence: without the retrieval loss, the encoder's standalone retrieval capability is severely degraded, rendering it unusable for the first-stage retrieval in an end-to-end pipeline. The multi-task design thus serves a strategic purpose: it preserves the encoder's dual utility as both a retrieval model and a component of the reranking framework, without compromising reranking quality.

\subsection{Impact of Input Document Order}

A known vulnerability of LLM-based listwise rerankers is sensitivity to the input ordering of candidate passages~\cite{liu2024lost}. Table~\ref{tab:order} examines ResRank's behavior under three input orderings: the original BM25-retrieved order, the inverse (reversed) order, and a random permutation.

\begin{table*}[t]
\centering
\caption{Impact of input document ordering on ResRank's ranking performance (nDCG@10).}
\label{tab:order}
\resizebox{\textwidth}{!}{
\begin{tabular}{l|cccccccc|c|cc}
\toprule
\textbf{Input Order} & \textbf{Covid} & \textbf{NFCorpus} & \textbf{Touche} & \textbf{DBPedia} & \textbf{SciFact} & \textbf{Signal} & \textbf{News} & \textbf{Robust} & \textbf{BEIR Avg.} & \textbf{DL19} & \textbf{DL20} \\
\midrule
Original & 0.8409 & \textbf{0.3973} & \textbf{0.3948} & \textbf{0.4583} & \textbf{0.7642} & \textbf{0.3527} & \textbf{0.5485} & \textbf{0.5964} & \textbf{0.5440} & \textbf{0.7359} & \textbf{0.6865} \\
Inverse & 0.8260 & 0.3829 & 0.2646 & 0.3782 & 0.6892 & 0.2318 & 0.5131 & 0.5672 & 0.4816 & 0.6822 & 0.6313 \\
Random & \textbf{0.8699} & 0.3856 & 0.3423 & 0.4003 & 0.7246 & 0.2769 & 0.5457 & 0.5856 & 0.5164 & 0.6993 & 0.6577 \\
\bottomrule
\end{tabular}
}
\end{table*}

The results reveal that ResRank exhibits sensitivity to input ordering, consistent with the positional biases inherent in causal language models. The original (BM25-descending) ordering yields the best performance, as it places more relevant passages earlier in the sequence where they benefit from stronger attention signals. Inverting the order degrades average BEIR performance by over 6 points, with particularly severe drops on Touche and Signal. Random ordering falls between the two extremes.

This ordering effect arises because the causal attention mechanism in the Reranker-LLM processes passage embeddings sequentially: later positions attend to all earlier positions, but not vice versa. When relevant passages are placed at the end (inverse order), they benefit from richer cross-passage context but the hidden states of earlier (less relevant) passages do not benefit from attending to later relevant passages, creating an imbalanced information flow. However, we note that this sensitivity is a characteristic shared by virtually all causal-attention-based listwise rerankers, and it is naturally mitigated in practice because first-stage retrievers already produce roughly relevance-ordered candidate lists. The development of techniques to further reduce positional sensitivity, such as bidirectional attention mechanisms or position-aware training augmentation, remains an interesting direction for future work.

% ============================================================
%                     5. CONCLUSION
% ============================================================
\section{Conclusion}

We have presented ResRank, a unified framework for efficient and effective LLM-based listwise passage reranking that addresses the fundamental efficiency bottlenecks of existing approaches through three synergistic innovations: residual passage compression that reduces each passage to a single contextualized embedding, cosine-similarity-based scoring that eliminates autoregressive decoding entirely, and dual-stage multi-task end-to-end joint training that aligns the learning objectives of retrieval and reranking. Experimental results demonstrate that ResRank achieves competitive or superior ranking effectiveness compared to strong baselines while requiring zero generated tokens and processing only one token per passage, yielding a fundamentally better balance between effectiveness and efficiency. End-to-end experiments further confirm that the jointly trained encoder and reranker exhibit aligned behavior, enabling a unified retrieval-reranking pipeline that achieves the best overall performance when complemented by sparse retrieval signals.

Looking ahead, several directions merit further investigation. First, exploring methods to mitigate the input order sensitivity inherent in causal attention, such as bidirectional attention or order-agnostic training strategies, could further improve robustness. Second, scaling ResRank to larger backbone models and longer candidate lists may reveal additional benefits. Third, extending the framework to multi-modal retrieval scenarios---where passages may contain images, tables, or structured data---represents a natural application of the compression-based architecture. Finally, integrating reinforcement learning from human feedback~\cite{zeng2025xhsrl} or preference optimization~\cite{deng2025onerec} into the training pipeline may further enhance ranking quality, particularly for domain-specific applications.

% ============================================================
%                     APPENDIX
% ============================================================
\appendix
\section{Annotation Prompt}
\label{sec:annotation_prompt}

To ensure high-quality training data, we employ the following prompt template to re-annotate relevance labels using Qwen3-Max. Given a query and $N$ candidate passages, the model is asked to both rank the passages and assign fine-grained relevance labels on a 4-point scale.

\begin{lstlisting}
I will provide you with {N} passages, each
indicated by a numerical identifier [].
Rank the passages based on their relevance
to the search query: {query}.

Documents:
[1] {document 1}
[2] {document 2}
...
[N] {document N}

Search Query: {query}

Rank the {N} passages above based on their
relevance to the search query. All the
passages should be included and listed using
identifiers, in descending order of relevance.

Additionally, assign a relevance label to
each document using the following 4-point
scale:
- 3: Highly Relevant (Directly answers
     the query)
- 2: Relevant (Contains useful related
     information)
- 1: Slightly Relevant (Mentions keywords
     but low utility)
- 0: Irrelevant (Unrelated to the query)

Please output the result strictly in the
following JSON format. Do not include
anything else. The "ranking" field should be
a string representing the order, and "labels"
should be a dictionary mapping each document
identifier to its score.

{
 "ranking": "[4] > [2] > [1] > [3] > ...",
 "labels": {"[1]": 1, "[2]": 3, ...}
}
\end{lstlisting}

% ============================================================
%                     REFERENCES
% ============================================================
\bibliographystyle{IEEEtran}

\end{document}